\documentclass[11pt]{article}
\usepackage{graphicx}

\def \lim {{\rm lim}}

\def \fine {{\rm fine}}
\def \coarse {{\rm coarse}}
\def \model {{\rm model}}

\def \CPU {{\rm CPU}}
\def \task {{\rm task}}

\def \data {{\rm data}}
\def \none {{\rm empty}}
\def \filled {{\rm filled}}

\def \cut {{\rm cut}}

\def \xb {{\bf x}}

\def \thetab {{\bf \theta}}
\def \thetabt {{\tilde{\thetab}}}

\def \Lc {{\Lambda}}

\def \sL {{\cal L}}

\def \BDM {\begin{displaymath}}
\def \EDM {\end{displaymath}}
\def \BEQ {\begin{equation}}
\def \EEQ {\end{equation}}
\def \BEQA {\begin{eqnarray}}
\def \EEQA {\end{eqnarray}}
\def \NN {\nonumber}
\def \BL {\begin{list}}
\def \EL {\end{list}}
\def \BENUM {\begin{enumerate}}
\def \EENUM {\end{enumerate}}
\def \BITEM {\begin{itemize}}
\def \EITEM {\end{itemize}}
\def \BARR {\begin{array}}
\def \EARR {\end{array}}

\begin{document}


\title{Cluster Detection in Astronomical Databases:
the Adaptive Matched Filter Algorithm and Implementation
\thanks{
This research was supported by  NSF grants AST93-15368 and AST 96-16901
and the Princeton University Research Board.
}
\thanks{
This work is sponsored by DARPA, under Air Force Contract
F19628-95-C-0002.  Opinions, interpretations, conclusions and
recommendations are those of the author and are not necessarily endorsed
by the United States Air Force. 
}}

\author{Jeremy Kepner$^{a,b}$ and Rita Kim$^a$ \\
$^a$Peyton Hall, Princeton University, Princeton, NJ 08544\\
$^b$Current Address: MIT Lincoln Laboratory,
    244 Wood St., Lexington, MA 02420
}
\maketitle

\begin{abstract}

  Clusters of galaxies are the most massive objects in the Universe and
mapping their location is an important astronomical problem. This paper
describes an algorithm (based on statistical signal processing methods),
a software architecture (based on a hybrid layered approach) and a
parallelization scheme (based on a client/server model) for finding
clusters of galaxies in large astronomical databases. The Adaptive
Matched Filter (AMF) algorithm presented here identifies clusters by
finding the peaks in a cluster likelihood map generated by convolving a
galaxy survey with a filter based on a cluster model and a background
model. The method has proved successful in identifying clusters in real
and simulated data.  The implementation is flexible and readily executed
in parallel on a network of workstations.

\end{abstract}

\section{Introduction}



  Clusters of galaxies are the largest objects known to humans (see
Figure \ref{fig:colley_cluster}).  They are the ``mountains'' of the
Cosmos, and like terrestial mountains they lie in great ranges that
define the cosmic ``continents'' and ``oceans'' (see Figure
\ref{fig:ncsa_sheets}). Mapping clusters of galaxies is very much
akin to surveying our own world and allows us to understand the
creation, evolution and eventual fate of our Universe \cite{Bahcall88}.

  The process by which astronomers detect clusters of galaxies
begins with assembling large images of the sky,
which are the result of hundreds of nights of observing through a
telescope. These pictures are analyzed to produce a database of
galaxies $X = \{ \xb_i : i = 1, \ldots, N_X \}$, where $N_X \sim 10^8$.
Each record $\xb_i \in X$ consists of a position on the sky,
brightness measurements in one or more bands and possibly hundreds of
additional measurements describing the shape and composition of the
galaxy.

  Clusters are local density peaks in the three dimensional distribution
of galaxies across the Universe.  In 3D data, clusters are easy to
detect.  Unfortunately, the majority of distances to individual galaxies
are not known and can only be inferred statistically from empirical
models of their brightness.  Thus, it is difficult to differentiate
small nearby clusters from large far away clusters. The goal of cluster
detection and estimation is to create a catalog, $\Theta$, consisting of
thousands of clusters $\Theta = \{ \thetab_i : i = 1, \ldots, N_\Theta
\}$, where $N_\Theta \sim 10^4$.  Each cluster in this list,
$\thetab_i$, consists of a position on the sky, a distance estimate, a
size estimate and perhaps additional estimated properties of the
cluster.


  The first catalog of galaxy clusters was compiled by Abell
\cite{Abell58}, and has proved extremely useful to astronomers over the
past four decades.  Abell's catalog was created by visually inspecting
hundreds of photographic plates taken from the first Palomar Observatory
Sky Survey (POSS). Modern galaxy databases are too large for such
methods to be used today.  Subsequent efforts to detect clusters have
relied on Matched Filter techniques taken from statistical signal
processing (e.g., \cite{Lumsden92}, \cite{Dalton94}, \cite{Postman96},
\cite{Kawasaki97} and \cite{Bramel00}.  These methods have a strong
mathematical foundation, but require extensive prior information and are
often computationally prohibitive as they test every possible location
in the domain of the cluster space $\Omega_\Theta$ for the presence of a
cluster. The Adaptive Matched Filter \cite{Kepner99} that is described
later in this paper is a variation on the Matched Filter that uses a
hierarchical set of filters, as well as software coding and parallel
computing techniques that address some of the Matched Filter's
drawbacks. The AMF is adaptive in two ways. First, the AMF uses a two
step approach that first applies a coarse filter to find the clusters
and then a fine filter to provide more precise estimates of the distance
and size of each cluster. Second, the AMF uses the location of the data
points as a ``naturally'' adaptive grid to ensure sufficient spatial
resolution.

  A variety of other techniques have also been applied to the cluster
finding problem.  The compact nature of clusters make Wavelet based
signal processing approaches an appealing alternative \cite{Fang97}
and \cite{Fadda97}. Geometric approaches such as Voronoi tessellation
\cite{Ramella98} have also been used.  In this method each $\xb_i$ is
the seed for the tessellation.  Clusters are then found by computing the
volume of each tessel and selecting the points with the smallest
volume, which presumably have the highest density.  Such geometric methods
have the advantage that they require very little prior information.

  The Matched Filter, Adaptive Matched Filter, Wavelet and Voronoi
Tessellation approaches all use the three to five high affinity
dimensions of $X$ (i.e., angular position and brightness measurements).
These dimensions are continuous real variables that lend themselves to
Euclidean distance metrics.  Working in these lower dimensions allows
more compute intensive techniques which are necessary to de-project
clusters from the observed data domain $\Omega_X$ to the desired
underlying domain in which clusters exist $\Omega_\Theta$, i.e., angular
position, {\it distance} and size. More recently, there has been
interest in exploiting the low affinity dimensions that are
also available in galaxy databases \cite{Djorgovski97,Gal99}
to enhance detection.  As the understanding of these
methods increases, the exploitation of many dimensions should be
possible using advanced datamining techniques (see e.g. \cite{Fayyad96}
and \cite{Dasarathy99} and references therein). These methods have
enormous potential for detecting new clusters and possibly separating
them into distinct groups thus revealing new classes of galaxy
clusters.


  The rest of this paper presents in greater detail the AMF algorithm,
its implementation and results.  In section two a detailed derivation
of the AMF is given.  The derivation is meant to be sufficiently
general that it can lend itself to other types of databases. Section
three presents the implementation of the AMF using a layered software
architecture and a client/server parallelization model.  Again, these
methods are not limited to the specific problem presented here and are
applicable to a variety areas. In section four the results of applying
the AMF on simulated and real data are discussed.  Finally, section
five gives the summary and conclusions.

\section{Adaptive Matched Filter Algorithm}

   Matched filter techniques are widely used in statistical signal
processing.  The idea is to convolve the data with a model of the
desired signal.  In many instances this can be shown to be optimal
detection method in the least squares sense.  Applying matched filter
techniques to point set data is less common but has become the standard
method for detecting cluster of galaxies.  The advantage of matched
filter techniques is that they are mathematically rigorous, provide
well defined selection criteria and produce few false detections.

  The Adaptive Matched Filter \cite{Kepner99} enhances the matched
filter method by creating a pair of filters (each correct under own its
assumptions) which can be used to trade off computational complexity
versus sensitivity.  The filters are derived by computing the
likelihood a cluster exists at a particular point $\thetab \in \Omega_\Theta$
given the data $X$. Various likelihood functions can be derived; the
differences are due to the additional assumptions that are made about
the distribution of the data.  This section gives the mathematical
derivation of the two likelihood functions used in the AMF:
$\sL_\coarse$ and $\sL_\fine$.  Both derivations are conceptually based
on virtually binning the data, but make different assumptions about the
distribution of points in the virtual bins.

  Imagine dividing up the data domain $\Omega_X$  into bins.  We
assign to each bin a unique index $j$.  The expected number of data
points in bin $j$ given that their is a cluster at $\thetab$  is denoted
$n_\model^j(\thetab)$.  The number of data points actually found in bin $j$
is $n_\data^j$.  In general, the probability of finding $n_\data^j$
points in cell $j$ is given by a Poisson distribution
  \BEQ
        P_j(\thetab) = \frac{(n_\model^j(\thetab))^{n_\data^j} e^{-n_\model^j(\thetab)}}
                   {       n_\data^j ! }
  \EEQ
The likelihood of the data given the model is computed from the sum 
of the logs of the individual probabilities
  \BEQ
       \sL = \sum_j \ln P_j(\thetab) .
  \EEQ

\subsection{Coarse Grained $\sL$}
  If the virtual bins are made big enough that there are many galaxies
in each bin, then the probability distribution can be approximated by a
Gaussian
  \BEQ
        P_j(\thetab) = \frac{1}{\sqrt{2\pi n_\model^j}}
              \exp \left \{
                 - \frac{(n_\data^j - n_\model^j)^2}
                        { 2 n_\model^j}
                   \right \} ~ .
  \EEQ
Furthermore, let the model distribution consist of a background field
(that is independent of $\thetab$) and a cluster component (that depends
on $\thetab$)
  \BEQ
        n_\model^j(\thetab) = n_f^j + n_c^j(\thetab) ~ .
  \EEQ
If the field contribution is approximately
uniform and large enough to dominate the noise then
  \BEQ
        P_j(\thetab) = \frac{1}{\sqrt{2\pi n_f^j}}
              \exp \left \{
                   - \frac{(n_\data^j - n_\model^j)^2}{ 2 n_f^j}
                  \right \} ~ .
  \EEQ
Summing the logs of these
probabilities results in the following expression for the coarse
likelihood
  \BEQA
       \sL_\coarse(\thetab)
       & = & \sum_j \ln P_j \NN \\
       & = & -\frac{1}{2} \sum_j \ln 2 \pi n_f^j
         -   \frac{1}{2} \sum_j 
               \frac{(n_\data^j - n_\model^j)^2}{ n_f^j}
  \EEQA
The first term is independent $\thetab$
and can be dropped. In addition, if the bins can also be made
sufficiently small, then the sum over all the bins can be replaced by an
integral
  \BEQ
        \sL_\coarse(\thetab) = -\int_{\Omega_X}
            \frac{(n_\data(\xb) - n_\model(\xb;\thetab))^2}
                 { n_f(\xb)} d\xb ~ ,
  \EEQ
where $n_\model^j(\thetab) = n_\model(\xb_j;\thetab)d\xb$ and  $n_\data(\xb)$
is a sum of Dirac delta functions corresponding to the locations of the
points $\xb_i$. Expanding the squared term and replacing $n_\model$
with $n_f + n_c(\thetab)$ yields
  \BEQ
        \sL_\coarse(\thetab) = -\int_{\Omega_X}
            \frac{n_\data^2 - 2 n_\data n_f - 2 n_\data n_c
                  + n_f^2 + 2 n_f n_c + n_c^2}
                 { n_f} d\xb ~ .
  \EEQ
The above expression can be simplified by setting $\delta = n_c/n_f$,
dropping all expressions that are independent of $\thetab$, and noting that
$\int n_c(\xb;\thetab) d\xb$ is small compared to the other terms, which leaves
  \BEQ
        \sL_\coarse(\thetab) = 2 \sum_{i=1}^{N_X} \delta(\xb_i;\thetab)
                    - \int_{\Omega_X}
                         \delta(\xb;\thetab) n_c(\xb;\thetab) d\xb ~ .
  \EEQ

\subsection{Fine Grained $\sL$}

  If the virtual bins are chosen to be sufficiently small that no bin
contains more than one galaxy, then the calculation of $\sL$ can be
significantly simplified because there are only two probabilities that
need to be computed. The probability of the empty bins
  \BEQ
        P_\none  =  e^{-n_\model^j} 
  \EEQ
and the probability of the filled bins
  \BEQ
        P_\filled  = n_\model^j e^{-n_\model^j} .
  \EEQ
The sum of the log of the probabilities is then
  \BEQA 
        \sL_\fine &=& \sum_\none \ln P_\none
                   +  \sum_\filled \ln P_\filled \NN \\
                  &=& - \sum_\none n_\model^j
                      - \sum_\filled n_\model^j
                      + \sum_\filled \ln n_\model^j
  \label{eq:finesum}
  \EEQA

  By definition summing over all the empty bins and all the filled bins
is the same as summing over all the bins.  Thus, the first two terms in
equation (\ref{eq:finesum}) are just the total number of points
predicted by the model
  \BEQA
        \sum_{\rm all~bins} n_\model^j
        &=& \int_{\Omega_X} n_\model(\xb;\thetab) d\xb \NN \\
        &=& N_\model(\thetab) = N_f + N_c(\thetab) ~ .
  \EEQA
$N_f$ and $N_c$ are the total number of field points and cluster points
one expects to see inside $\Omega_X$; they can be computed by
integrating $n_f$ and  $n_c$
  \BEQA
       N_f &=& \int_{\Omega_X} n_f(\xb) d\xb ~ , \NN \\
       N_c(\thetab)
       &=& \int_{\Omega_X} n_c(\xb;\thetab) d\xb ~ .
  \EEQA

  Because we retain complete freedom to locate the bins wherever we like, we
can center all the filled bins on the points $\xb_i$, in which case the third
term in equation (\ref{eq:finesum}) becomes
  \BEQ
        \sum_\filled \ln n_\model^j \rightarrow
        \sum_{i=1}^{N_X} \ln n_\model(\xb_i;\thetab) =
        \sum_{i=1}^{N_X} \ln[n_f(\xb_i) + n_c(\xb_i;\thetab)] ~ ,
  \EEQ
and the sum is now carried out over all the points instead of all the
filled bins. Combining these results we can now write the likelihood in
terms that are readily computable from the model and the database
  \BEQ
        \sL_\fine(\thetab) = - N_f - N_c + \sum_i \ln[n_f(\xb_i) + n_c(\xb_i;\thetab)] ~ .
  \EEQ
Subtitutuing $\delta = n_c/n_f$ and dropping terms that are independent of
$\thetab$ gives:
  \BEQ
        \sL_\fine(\thetab) = - N_c + \sum_i \ln[1 + \delta(\xb_i;\thetab)] ~ .
  \EEQ

\subsection{Application}

  Both likelihood functions are applied to the data in a similar
manner. A set of $N_\Theta^{\rm test}$ test locations are chosen from
the cluster domain $\Omega_\Theta$. The likelihood functions are then
evaluated at each test location to produce a likelihood map.  The
clusters correspond to the peaks in this map that are above a specified
threshold. In full generality, producing the likelihood map would
require a O($N_X N_\Theta^{\rm test}$) function evaluations.  For a specific
dataset, the model functions $n_f(\xb)$ and $n_c(\xb;\thetab)$ are
constructed using prior empirical and theoretical knowledge of the data
(see \cite{Kepner99} for the specific functions). From these models
additional symmetries emerge which can be exploited to significantly
reduce the computations.  For example, galaxy clusters have a finite
angular size so at each test location only the small sub-set of
data points which are near the test location need to be considered.

  Another simplification comes from the fact that clusters have a shape
that is roughly independent of the total number of galaxies in the
cluster 
  \BEQ
        n_c(\xb;\thetab) \rightarrow \Lc n_c(\xb;\thetabt)
  \EEQ
where $\thetab = (\thetabt, \Lc)$, and $\Lambda$ parameterizes
the size of the cluster.  This simple modification allows the coarse
likelihood function to be re-written as
  \BEQ
        \sL_\coarse(\thetab) = 2 \Lc \sum_i \delta(\xb_i;\thetabt)
                    - \Lc^2 \int_{\Omega_X}
                         \delta(\xb;\thetabt) n_c(\xb;\thetabt) d\xb ~ ,
  \EEQ
which can now be solved for $\Lc$ by setting $\partial \sL / \partial \Lc = 0$
  \BEQ
        \Lc_\coarse(\thetabt) = \frac{
              \sum_i \delta(\xb_i;\thetabt)
           }{
              \int \delta(\xb;\thetabt) n_c(\xb;\thetabt) d\xb ~ .
          }
  \EEQ
Inserting this value back into the previous equation gives
  \BEQ
        \sL_\coarse(\thetab) = \Lc_\coarse(\thetabt) \sum_i \delta(\xb_i;\thetabt) ~.
  \EEQ
The result of the above simplification is the elimination of one
of the search dimensions, which results in a sizeable computational savings.

  The same simplification when applied to the fine likelihood function gives
  \BEQ
        \sL_\fine = - \Lambda_\fine  N_c + \sum_i \ln[1 + \Lambda_\fine \delta(\xb_i;\thetabt)]
  \EEQ
where $\Lc_\fine$ is computed by solving
  \BEQ  
       N_c = \sum_i \frac{\delta_i}{1 + \Lambda_\fine \delta(\xb_i;\thetabt)} ~ .
  \EEQ
While $\Lc_\coarse$ can be obtained directly from $\sL_\coarse$,
$\Lc_\fine$ can only be found by numerically finding the zero point of
the above equation. Furthermore, this equation does not lend
itself to standard derivative based solvers (e.g., Newton-Raphson) that
produce accurate solutions in only a few iterations.  Fortunately, the
solution can usually be bracketed in the range $0 < \Lc_\fine < 1000$,
thus obtaining a solution with an accuracy $\Delta \Lc \sim 1$ takes
$\log_2 (1000/1) = 10$ iterations using a bisection method.

\begin{figure}[tbh]
\centerline{\includegraphics[width=6.5in]{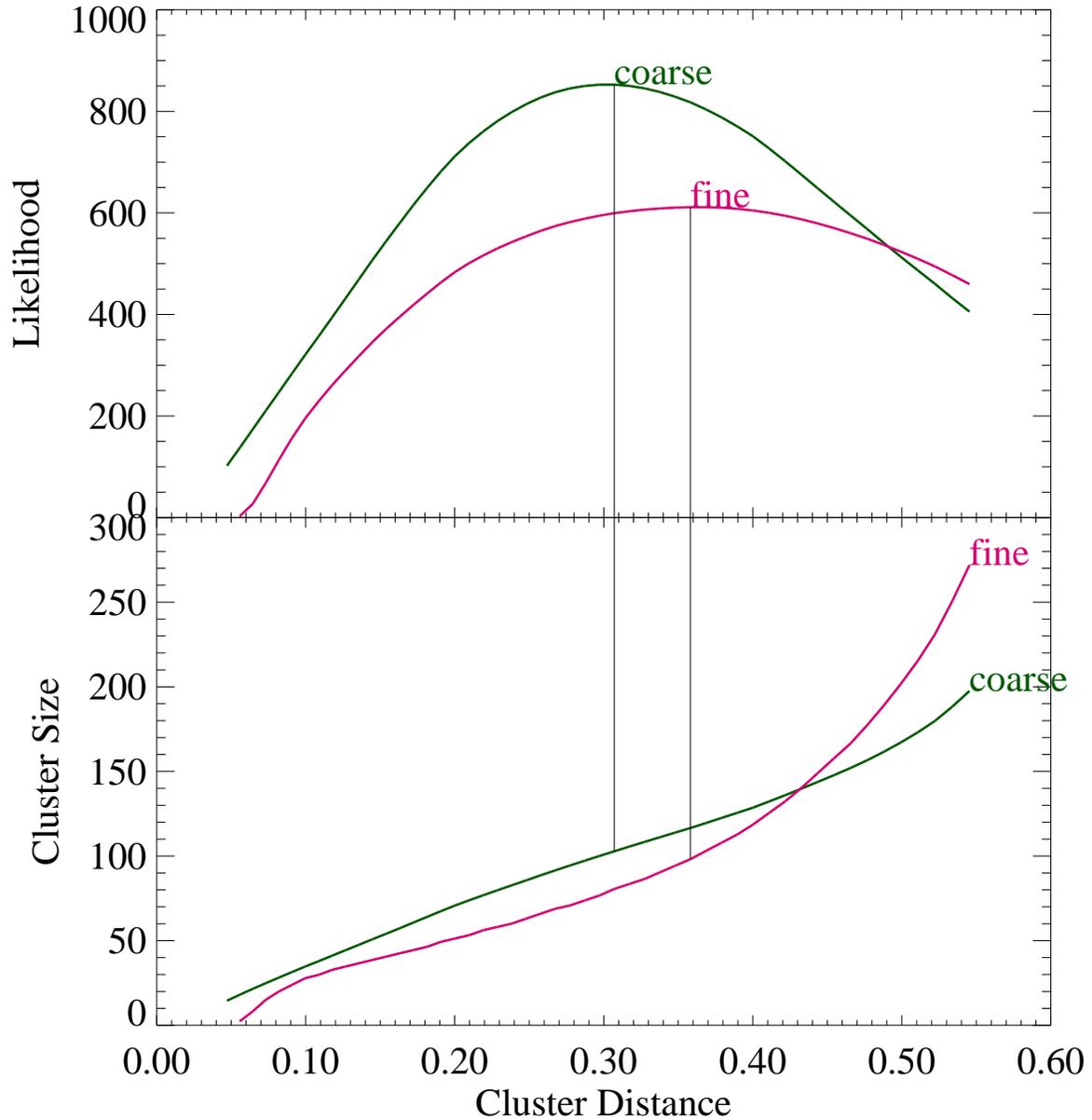}}
\caption{ {\bf Coarse and Fine Filters.}
  Plots of the of likelihood and cluster size ($\Lc$) as a function
of distance as computed from the coarse and fine matched filters.  The
input cluster has a true distance of 0.35 (4.5 billion light years) and
size of $\Lc = 100$. The coarse likelihood peaks at a distance of 0.31
and estimates the size to be $\Lc = 102$.  The fine likelihood peaks at
a distance of 0.35 and a estimates the size to be $\Lc = 98$.  In
general, the fine likelihood provides better distance and size
estimates but at approximately 10 times the computational cost.
}
\label{fig:coarse_vs_fine}
\end{figure}

  Both the coarse and fine likelihood functions are able to exploit the
specifics of the model to significantly reduce the number of test
locations that need to be evaluated. The coarse likelihood function
requires about 10 times less work to evaluate than the fine likelihood.
Unfortunately, the underlying assumptions used in the derivation of the
coarse likelihood function are not as accurate as those used to derive
the fine likelihood.  Thus, while the coarse likelihood is faster, the
fine likelihood is more accurate (see Figure~\ref{fig:coarse_vs_fine}).
The AMF addresses this issue by using both likelihood functions in a
two stage approach.  First, the coarse likelihood function is applied
and then the fine likelihood function is used on the peaks found in the
coarse map.

  Using both filters sequentially not only produces the best estimate of
the cluster locations, it has the added benefit of providing two
quasi-independent sets of values for each cluster.  This provides a
helpful consistency check because the coarse and fine filter react
differently at the detection limit.   The coarse likelihood tends to
assign weak detections to small nearby clusters, while the fine
likelihood makes these detections large, far away clusters. Thus, if
both likelihoods peak at similar size and distance estimates, then the
detections are probably real, but if the two likelihoods peak at
dramatically different values than the cluster is probably a false
detection.

\section{Implementation}

  The likelihood functions derived in the previous section represent the
core of the AMF cluster detection scheme. Both likelihood functions
begin with picking a grid of test locations.  The most straightforward
method is to use a regularly spaced grid over $\Omega_\Theta$. Recall
that each point in $\Theta$ consists of an angular position, a distance
and a cluster size and that each point in $X$ consists of an angular
position and a brightness.  As shown in the previous section, the size
can be determined without searching, and a regular grid in distance will
work reasonably well provided the steps are sufficiently small (see
Figure~\ref{fig:coarse_vs_fine}). A regular grid in angle has the
difficulty of making the grid too big in dense regions and too fine in
sparse regions (i.e., it is unnecessary to search for clusters where
there is no data).  A more optimal set of test locations is to use the
angular positions of the data $\xb_i$, which``naturally'' provides an
adpative resolution.

\subsection{Peak Finding}

  Finding peaks in a 3D regularly gridded map is straightforward.
Finding the peaks in the irregularly gridded map is more
difficult.  There are several possible approaches. We present a simple
method which is sufficient for selecting individual clusters.  More
sophisticated methods will be necessary in order to find small clusters
that are close to large clusters.

  As a first step we eliminate all low likelihood points $\sL_\coarse^i
< \sL_\cut$, where $\sL_\cut$ is the nominal detection limit, which is
independent of richness or redshift.  $\sL_\cut$ can be estimated from
the distribution of the $\sL_\coarse^i$ values.  Step two consists of
finding the largest value of $\sL_\coarse^i$, which is by definition
the first and largest cluster $\thetab_1$.  The third step is to eliminate
all test points that are within a certain radius of the cluster. 
Repeating steps two and three until there are no points left results in
a complete cluster list $\Theta$. A different scheme would be to connect the
irregularly gridded points in a Voronoi tessellation \cite{Ramella98}
from which local maxima could be obtained in the same manner as on a
grid.

\subsection{Software Architecture}

  Implementation of the AMF cluster selection consists of four steps:
(1) reading the database and the model parameter files, (2) computing
$\sL_\coarse$ over the entire database, (3) finding clusters by
identifying peaks in the $\sL_\coarse$ map, and (4) evaluating
$\sL_\fine$ and obtaining a more precise determination of each
cluster's size and distance.

\begin{figure}[tbh]
\centerline{\includegraphics[width=6.5in]{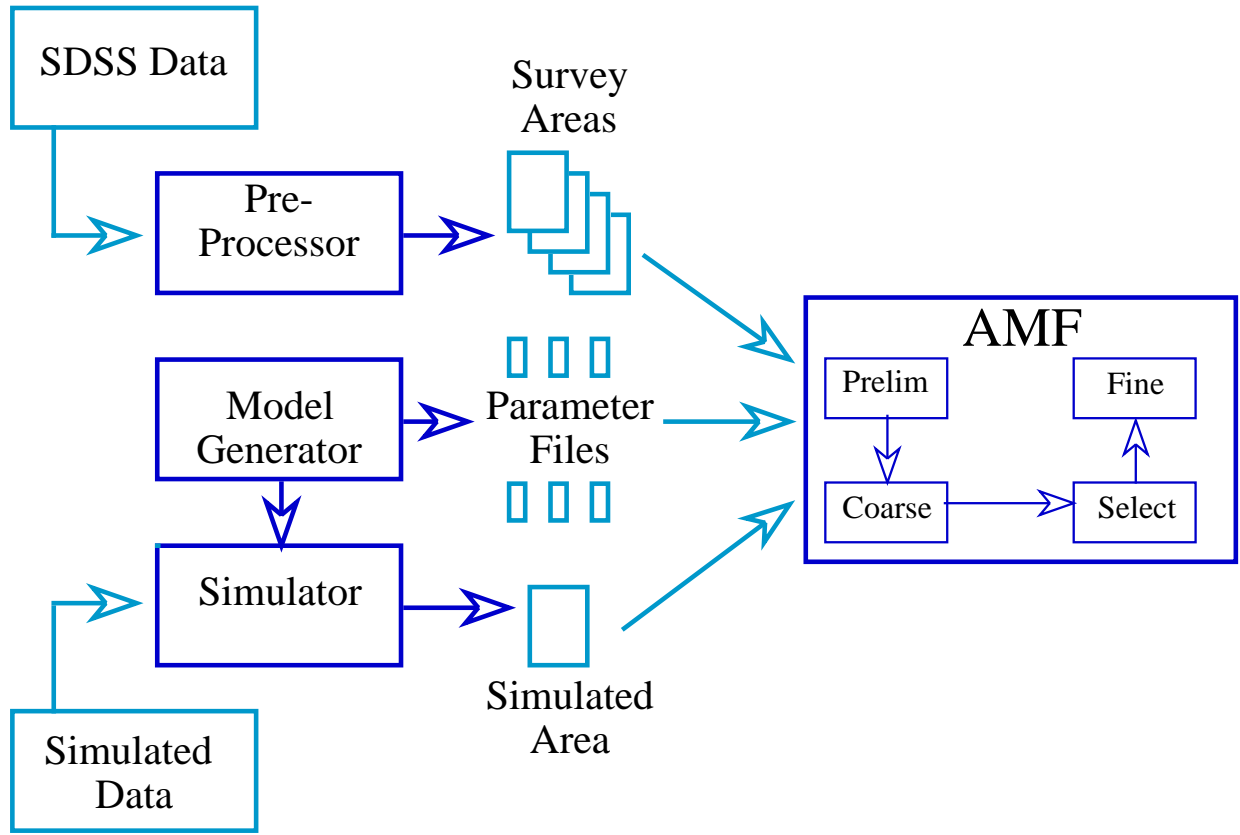}}
\caption{ {\bf Software Pipeline.}
  Schematic of the AMF software processing pipeline.  The processing
can accept either real or simulated data.  The data is broken up into
survey areas and fed into the AMF.  The AMF does preliminary data
checking followed by the coarse filter. The results of the coarse
filter are used to select the locations of clusters which are then fed
into the fine filter to provide more accurate estimates.  The coarse
filter is the dominant step in terms of computing cost.
}
\label{fig:software_pipeline}
\end{figure}

  The architecture of this data processing pipeline is shown in
Figure~\ref{fig:software_pipeline}.  The software has been designed
so that it can accept both real and simulated data.  One of the
challenges of the AMF is organizing the software so that it can readily
accept new datasets and different parameter files.  Critical to
adapting to new data is the ability see into the system and observe
each step as it takes place. To address these issues the vast majority
of the code has been implemented in an interpreted language (IDL from
Research Systems, Inc.) which provides many mechanisms for reading in
files and for monitoring and visualizing output.  

\begin{figure}[tbh]
\centerline{\includegraphics[width=6.5in]{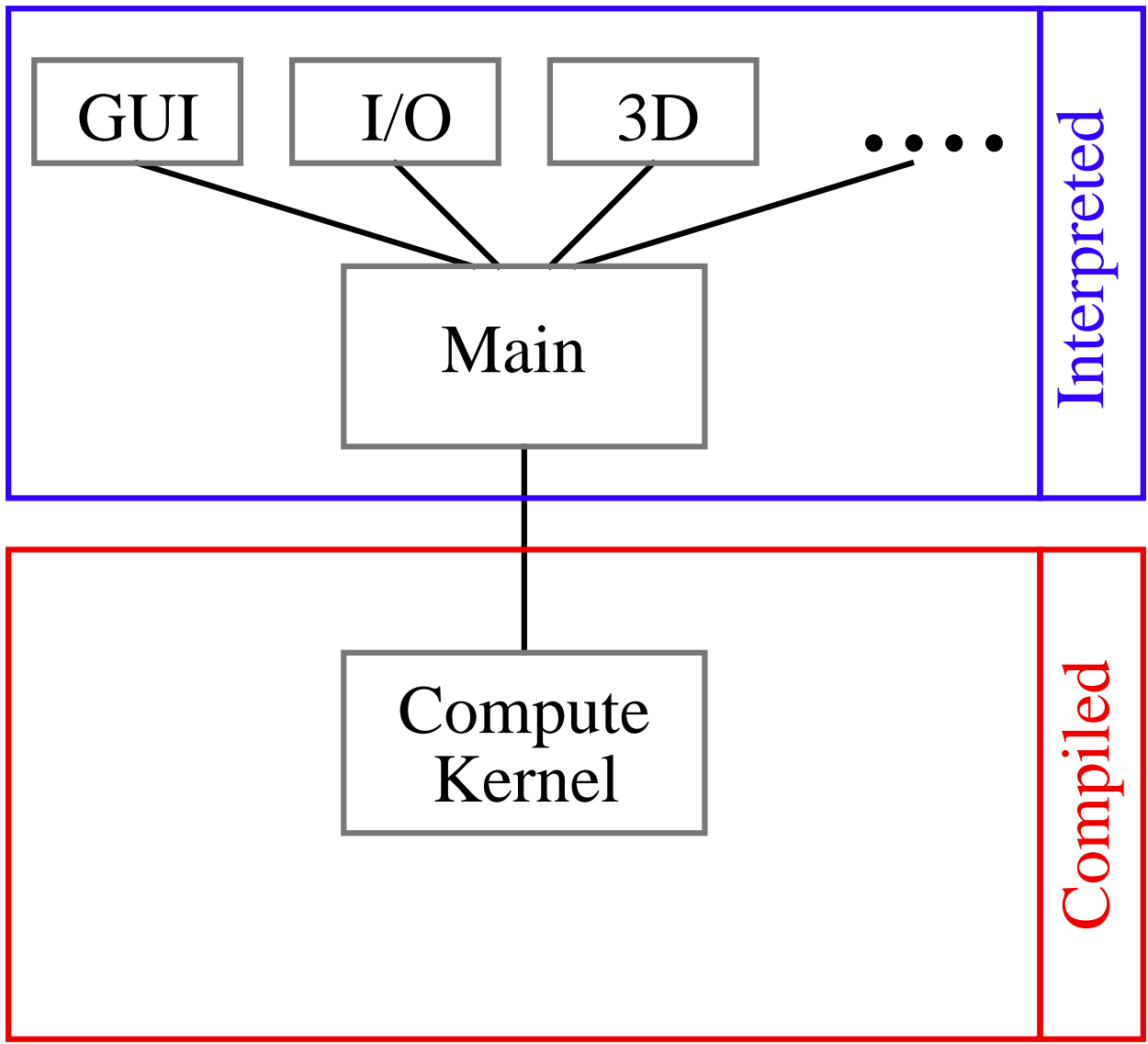}}   
\caption{ {\bf Single CPU Application Architecture.}
  Application architecture before implementation on an
Network-Of-Workstations. GUI and other ``high level'' operations are
written in the interpreted layer, which calls the compute kernel (the
coarse filter in the AMF application) written in a compiled language.
}
\label{fig:one_cpu_app_arch}
\end{figure}

  The computational driver of the application is the evaluation of
$\sL_\coarse$.  This function consists of a set of nested for loops
which do not lend themselves to the vector notation required to
get good performance in an interpreted language.  Thus, while
the interpreted code is used to set up the calculation, a compiled
C routine is called to compute the coarse likelihood function
(see Figure~\ref{fig:one_cpu_app_arch}).  In addition to giving
the superior compute performance of a compiled language, this
layered software approach also provides a mechanism for exploiting
parallel computing.

\subsection{Parallelization Scheme}

\begin{figure}[tbh]
\centerline{\includegraphics[width=6.5in]{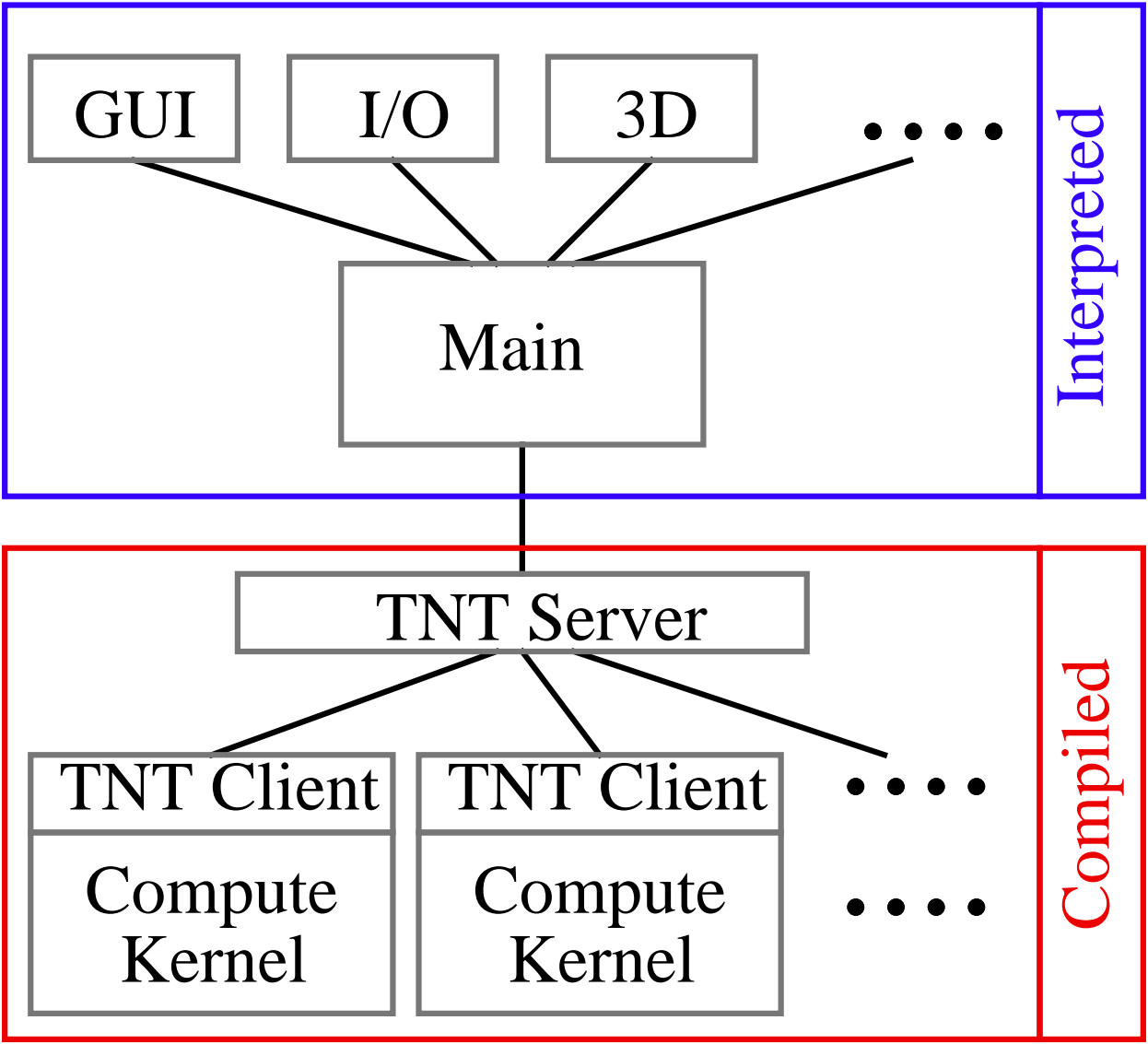}}
\caption{ {\bf NOW Application Architecture.}
  Application architecture after implementation on an NOW.
An additional ``TNT'' layer has been added to the compute kernel which
invokes and manages multiple copies of the compute kernel (coarse filter)
on a NOW.
}
\label{fig:mpnow_app_arch}
\end{figure}

  Computing the coarse likelihood map is a highly parallelizable
operation.  Each test point can be computed independently of the others
if all the necessary data is available.  There are a variety of ways to
take advantage of this scheme.  The one chosen here is a client/server
approach based on the The Next generation Taskbag (TNT) software
library \cite{Kepner00}.  TNT is a client-server based Applications
Programming Interface (API) for distributing and managing multiple
tasks on a Network-Of-Workstations (NOW).  TNT is a C based library
which can be used in any compiled program.  As such, it is possible to
insert the appropriate TNT calls into the compiled layer called by an
interpreted language (see Figure~\ref{fig:mpnow_app_arch}). 

\begin{figure}[tbh]
\centerline{\includegraphics[width=6.5in]{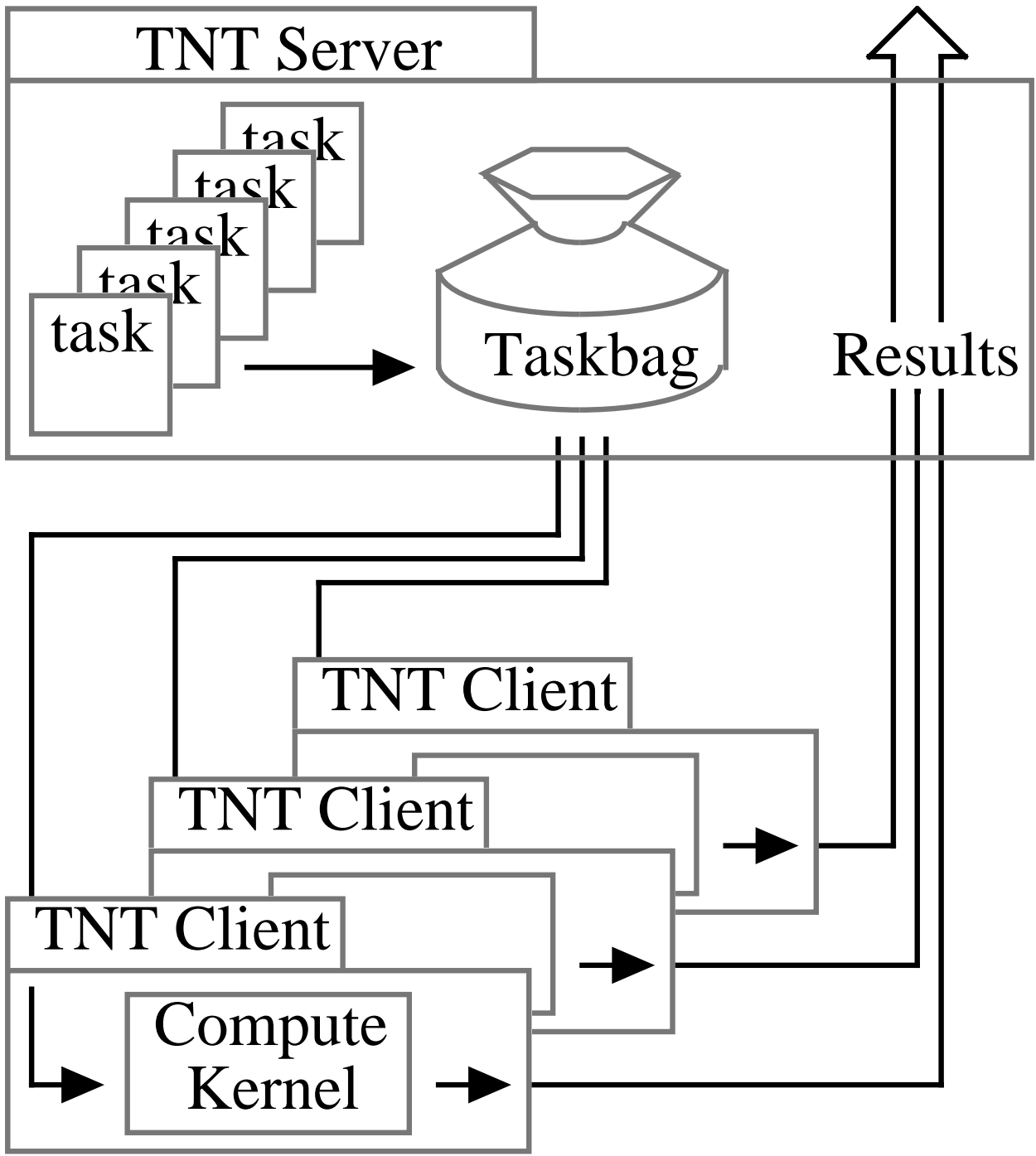}}
\caption{ {\bf TNT Application}
  A typical TNT application consists of a server with many clients,
communicating via TCP/IP.  The server: places tasks into Taskbag;
listens on a specific port for requests for tasks from clients;
dispatches tasks to requesting clients; accepts results from clients;
monitors status of clients and re-assigns tasks of dropped clients;
when all tasks are completed, returns results back to the main program.
The client(s) loop over the Taskbag is until it is empty.  On each
iteration a client will: send requests for work to server on a specific
port; read data sent by server over network; call compute kernel with
the data; send results of computation back to server over network.  For
the AMF, the tasks correspond to sub-sets of the test locations and the
compute kernel is the coarse likelihood evaluation.
}
\label{fig:tnt_app}
\end{figure}

  The operation of a typical TNT application is shown in
Figure~\ref{fig:tnt_app}.  The server creates a ``Taskbag'' of work for
clients.  The clients are then executed remotely on a number of
processors.  The clients connect with the server and request a task or
taskbag (a group of tasks).  When they have completed their tasks they
return the results back to the server and ask for more tasks. 

  The TNT library was developed on Linux (RedHat 5.0) and tested on
FreeBSD, NetBSD, and Solaris.  The entire library is written in C using
TCP sockets.  A server communicates with the clients using ports, which
allows simultaneous servers to be active and listening to different
port numbers. A server can call client functions and a client can call
server functions.  This enables the creation of hierarchies of servers.
For example, a ``root'' server can partition a large taskbag into many
sub-taskbags and distribute them to a collection of sub-servers.  These
sub-servers will then distribute tasks to the clients.  This allows for
a more efficient distribution of work across the cluster nodes. 

  For the AMF application, the interpreted language calls a C routine which
then sets up a server with tasks to be executed
(Figure~\ref{fig:tnt_app}).  In this case, each task is a sub-set of
all the test points.  Clients are then started on other computers (or
the same machine if it is multiple processor system). At startup each
client receives the database (or a portion of the database) from the
server.  After receiving the data, each client then asks the server for
a task to execute and returns the result.

  It is not possible to predict in advance how long a given task
is going to take because of the non-linearity of the algorithm and
because of heterogeneous capabilities and loads that may exist on the
NOW.  Fortunately, TNT is inherently load balancing in the sense that
when a client finishes a task it requests additional work. If there are
no tasks remaining then the client exits and frees up the processor. 
The processors that run faster will pick up more work and slower
processors will pick up less work.

\section{Results}

  The AMF has been extensively tested on simulated data to verify its
accuracy and robustness \cite{Kepner99}.   The AMF is currently being
applied to detect clusters of galaxies from the Sloan Digital Sky
Survey (SDSS) \cite{Kim00}.  The recent parallel implementation of the
AMF has significantly increased speed of the application.

\subsection{Tests on simulated data}

  In real data, neither the distances of the galaxies nor the position
and sizes of the clusters are known.  Tests on simulated data are the
only opportunity to check the detection algorithm in a well understood
environment. The test data consists of 72 simulated clusters with
different sizes and distances placed in a simulated field of randomly
distributed galaxies. The data
was constructed to be consistent with what is expected from SDSS.  The
clusters range in size and distance so as to span the full range of
expected clusters.  The test data covers an area of 10 square
degrees (1/1000 of the SDSS) and contains approximately 100,000
points.

\begin{figure}[tbh]
\centerline{\includegraphics[width=6.5in]{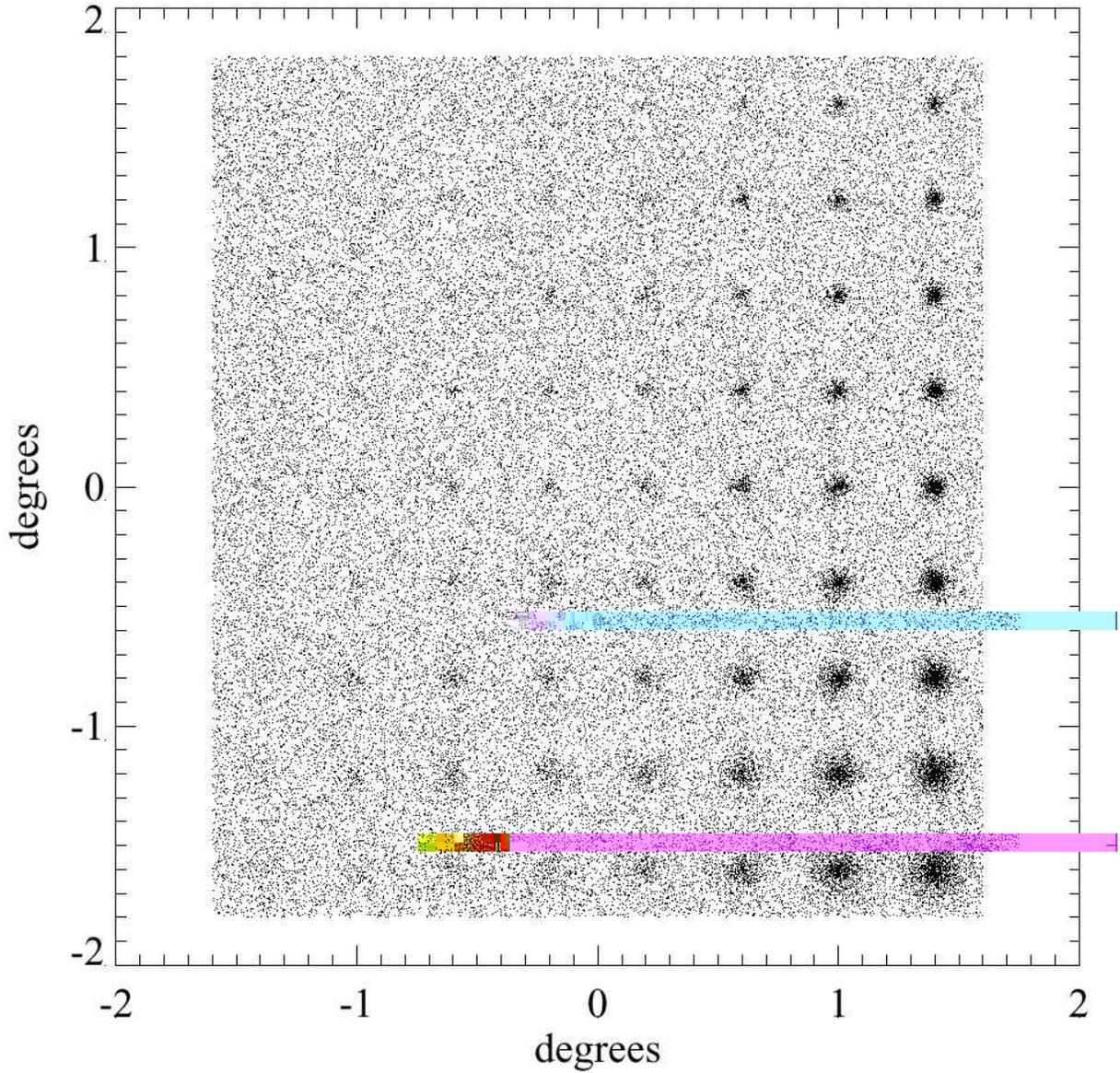}}
\caption{{\bf Simulated Data.}
  Angular positions of simulated data containing a uniform field and 72
clusters arranged in an 8 by 9 grid. Each column of clusters are the
same size while each row of clusters are at the same distance. From
left to right the sizes are $\Lambda =$ 10, 20, 30, 40, 50, 100, 200,
and 300.  From bottom to top the distances are $z =$ 0.1, 0.15, 0.2,
0.25, 0.3, 0.35, 0.4, 0.45, and 0.5.
}
\label{fig:simulated_data}
\end{figure}

  To facilitate the subsequent analysis and interpretation of the
results, the clusters were placed on an 8 by 9 grid. The cluster
centers were separated by 0.4 degrees. The distribution of all the
galaxies in angle is shown in Figure~\ref{fig:simulated_data}, where
each column of clusters are the same size while each row of clusters
are at the same distance. From left to right the sizes are $\Lambda =$
10, 20, 30, 40, 50, 100, 200, and 300.  From bottom to top the
distances are $z =$ 0.1, 0.15, 0.2, 0.25, 0.3, 0.35, 0.4, 0.45, and
0.5.

\begin{figure}[tbh]
\centerline{\includegraphics[width=6.5in]{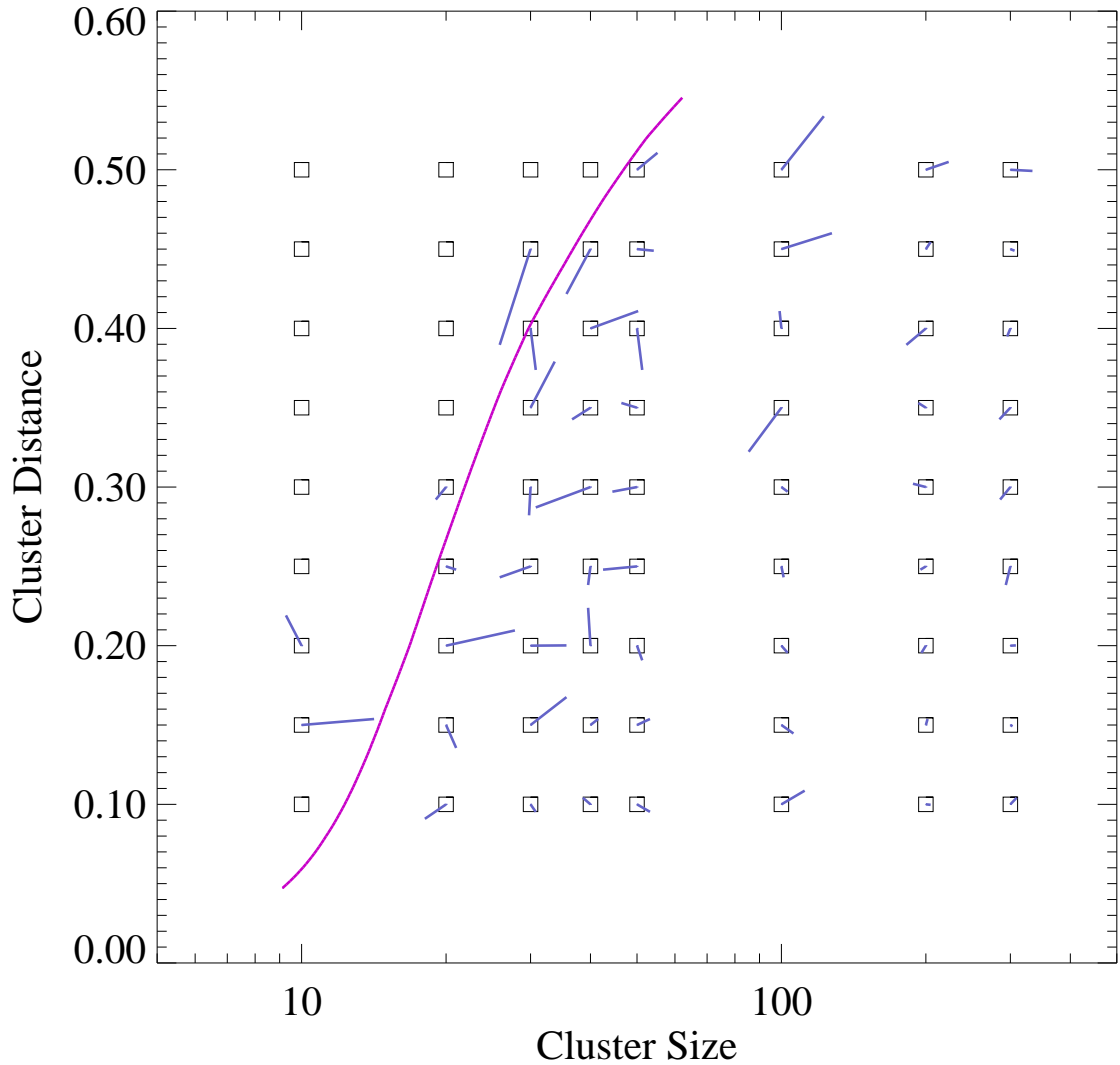}}
\caption{{\bf Results from Simulated Data.}
  The size and distance of each of the input clusters (boxes) with the
short lines indicating the corresponding values determined from the AMF
fine filter. The long curved line indicates the approximate detection
limit.  All the clusters above the detection limit (i.e. to the right of
the line) are found with no false detections.  As expected the size and
distance estimates are best for the largest and nearest clusters, while
clusters near the detection limit have poorer estimates.
}
\label{fig:simulation_results}
\end{figure}

  The simulated data were run through the AMF and all clusters above
the designated 5-$\sigma$ noise limited threshold were detected (with
no false detections).  The angular positions of all the detected
clusters were well within the expected range.  The estimated distance
and size of each cluster is shown in
Figure~\ref{fig:simulation_results}. As expected, the large and/or
nearby clusters are detected and measured more accurately than the
small, far away clusters.  These results indicate that the AMF can
detect and unbiasly estimate the location of clusters.

\subsection{Tests on SDSS data}
  
\begin{figure}[tbh]
\centerline{\includegraphics[width=6.5in]{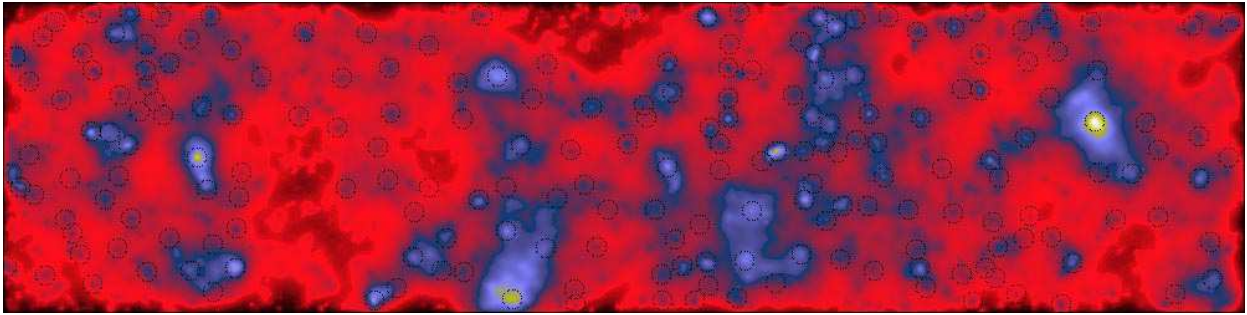}}
\caption{{\bf Likelihood Map of Sloan Data.}
  A strip of the Sloan Digital Sky Survey data (courtesy of SDSS
Collaboration) was processed through the AMF. This strip covers 30
square degrees and contains nearly 200,000 galaxies.   The image shows
the projected coarse likelihood map of this strip (red are low values,
blue higher, and green is highest). The the dotted circles denote cluster
detections.
}
\label{fig:sloan_results}
\end{figure}

  The Sloan Digital Sky Survey is a multi-decade, multi-institution
effort to take a million by million pixel composite image of the night
sky in five color bands.  Analysis of the image data is expected to
yield a database of 200,000,000 galaxies \cite{Szalay99}. Test data has
been taken since 1998 and been used to test all aspects of the SDSS
software including the AMF (see Figure~\ref{fig:sloan_results}). The
AMF has detected all previously known clusters in the SDSS test data it
has looked at.  In addition, the AMF performs at a level equal to or
better than the other algorithms with which it has been compared.
Detailed results and comparison of various cluster finding algorithms
are presented elsewhere \cite{Kim00}.

  Adapting the AMF to the SDSS was a sizeable effort. Without the
layered software approach it would have taken considerably longer as a
considerable amount of tuning was required to properly set all the
model parameters.

\subsection{Scalability results}

  A parallel implementation of the AMF is crucial for its application
to the full SDSS database.  Currently, the AMF requires 1-2 hours of
CPU time (450 MHz Pentium II) to process a 10 square degree field.
On a parallel NOW this can easily be sped up by a factor of 100,
which will make processing the entire 10,000 square degree SDSS
dataset feasible.

\begin{figure}[tbh]
\centerline{\includegraphics[width=6.5in]{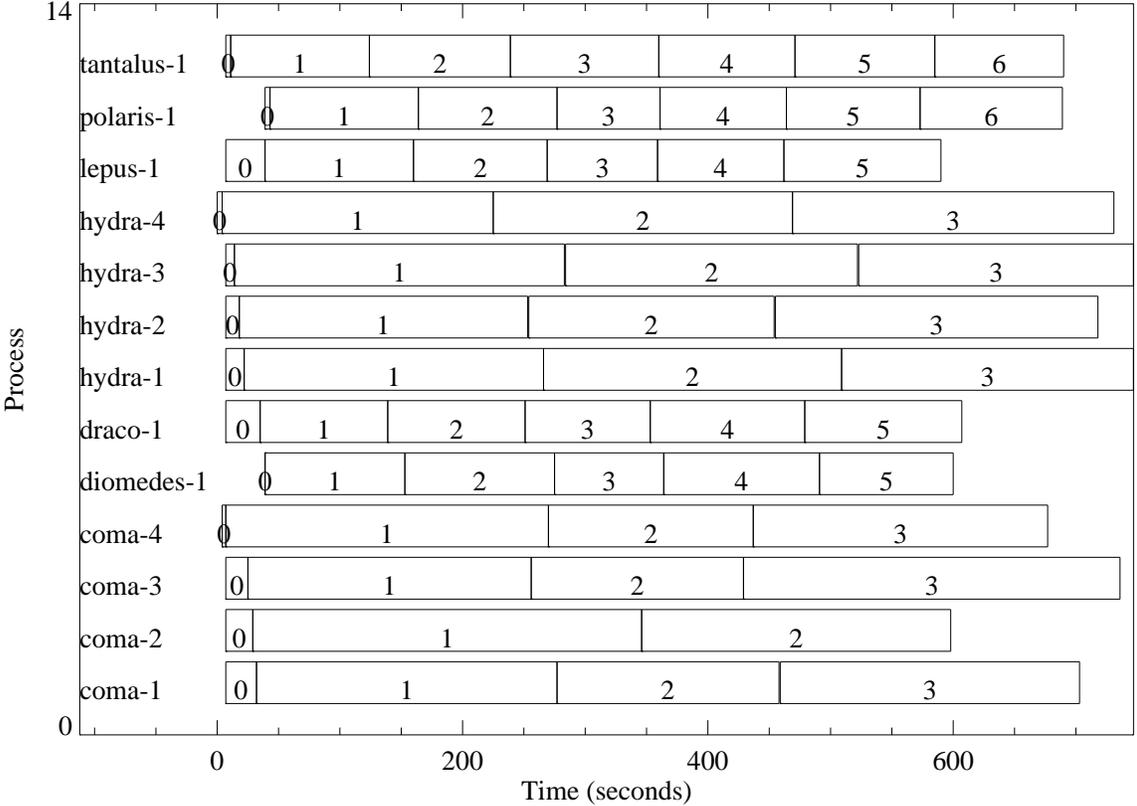}}
\caption{ {\bf Heterogeneous NOW (50 Tasks).}
  Task schedule for the coarse filter running on a heterogeneous NOW
consisting of seven single processor and two quad processor
workstations.  The number in each box shows how many tasks each
processor has completed.  Task 0 is the initial transmittal of the
data.  This calculation achieves a speedup of 7.8 out of 10.5 (74\%).
The barriers to full speedup are the initial transmission and the slack
time at the end due the task granularity.
}
\label{fig:timing_a}
\end{figure}

\begin{figure}[tbh]
\centerline{\includegraphics[width=6.5in]{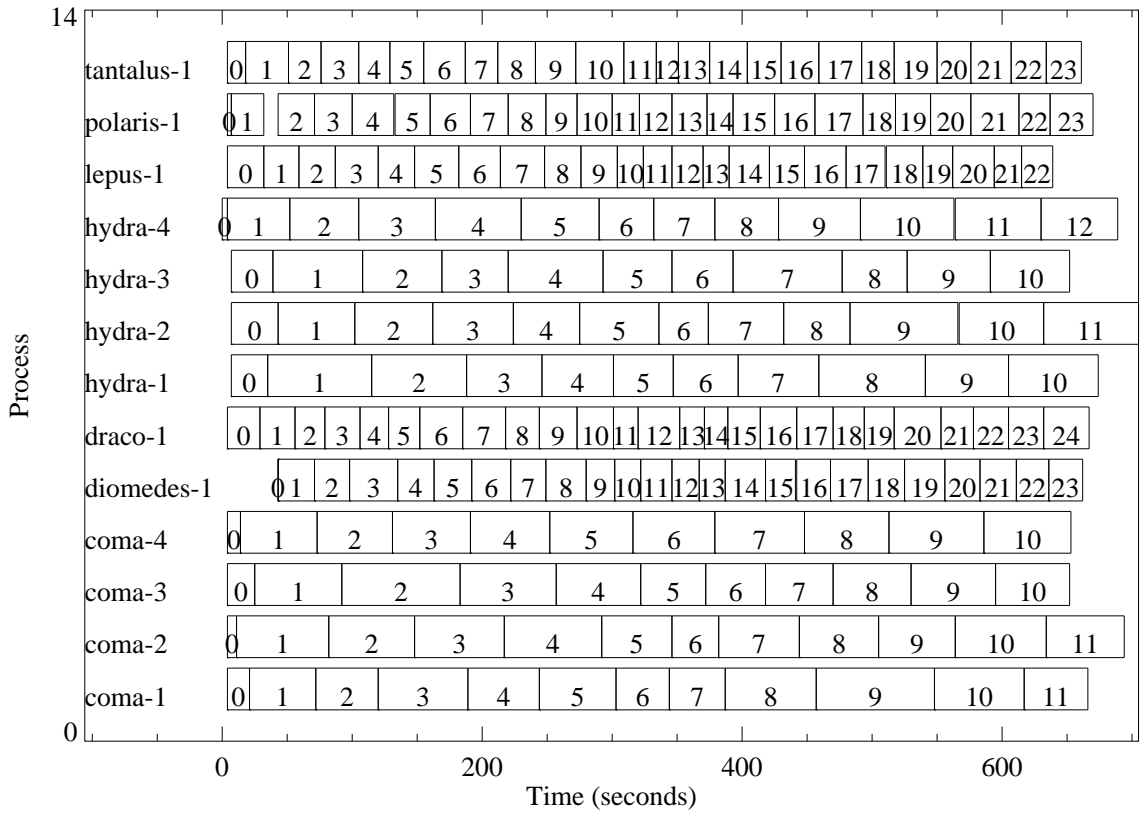}}
\caption{ {\bf Heterogeneous NOW (200 Tasks).}
  Same as Figure~\ref{fig:timing_a} except that the work has
been broken up into 200 tasks which reduces the slack time at
the end of the calculation resulting in a speedup of 8.4 out of
10.5 (80\%).
}
\label{fig:timing_b}
\end{figure}

\begin{figure}[tbh]
\centerline{\includegraphics[width=6.5in]{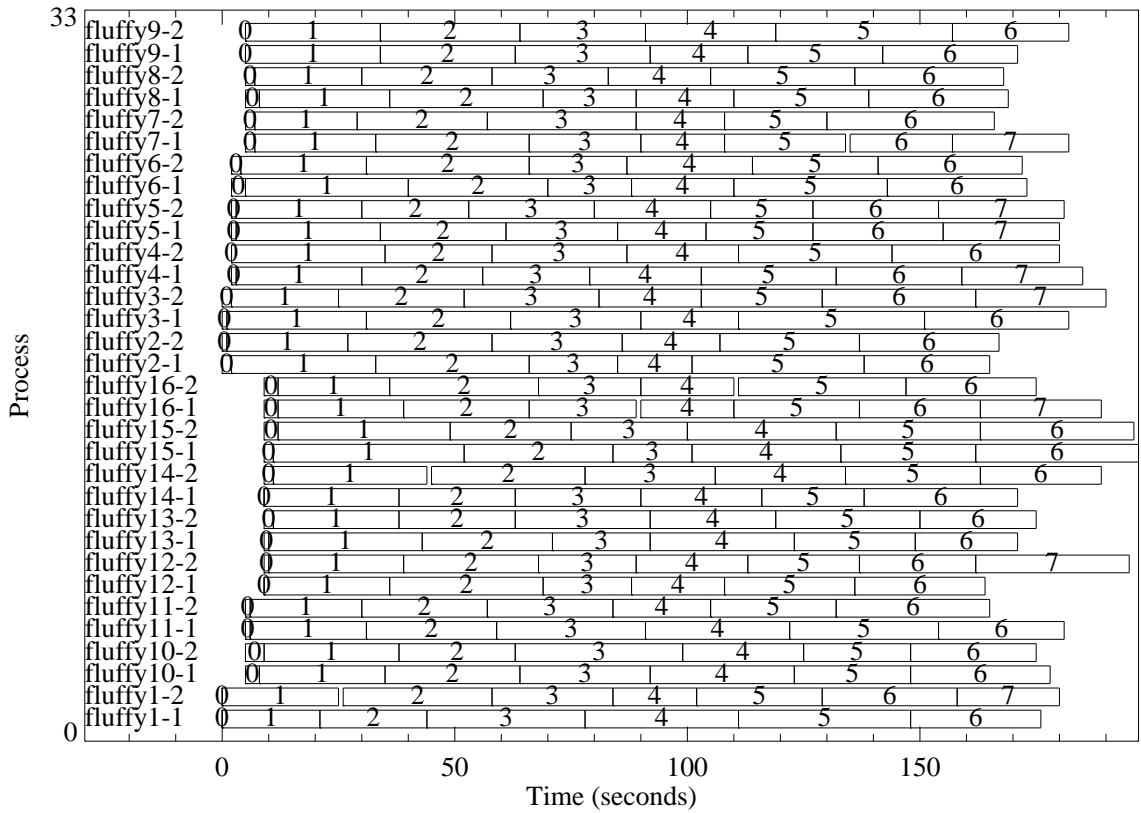}}
\caption{ {\bf Homogeneous NOW (200 Tasks).}
  Same as Figure~\ref{fig:timing_b} but run on sixteen identical dual
processor workstations with a higher performance network.  The better
network reduces the initial transmit time (Task 0) and results in a
speedup of 27.3 out of 32 (85\%).
}
\label{fig:timing_c}
\end{figure}

  The results of running the parallel implementation on a NOW are shown
in Table~1. These data show that the algorithm experiences good speedup
on both heterogeneous and homogeneous NOWs.  The primary bottlenecks to
perfect speedups are the initial sending of data and the granularity of
the tasks.  The impact of these can be seen in the execution schedules
shown in Figures~\ref{fig:timing_a}-\ref{fig:timing_c}.

  The total computation time consists of the time to do
the computation plus the slack time due to the granularity
of the tasks
  \BEQ
        T_{\rm comp} \propto \frac{N_X}{N_\CPU} + \frac{N_X}{N_\task} ~ ,
  \EEQ
where $N_\CPU$ is the number of processors used in the NOW and
$N_\task$ is the number of tasks the job was broken into. The total
time spent communicating is the time spent gathering the results plus
the time to send the initial data to each processor
  \BEQ
        T_{\rm comm} \propto \frac{N_X}{N_\CPU} + N_X N_\CPU ~ .
  \EEQ

  To achieve good scalability requires that the
computation-to-communication ratio stay high as $N_\CPU$ increases.  In
both cases the first term scales well while the second term doesn't. The
second (granularity) term in the computation time is due to the fact
that some processors will finish first and there will be no additional
tasks for them to complete.  This can be alleviated by simply dividing
the work up into sufficiently small tasks until this time no longer
becomes important.

  The second (startup) term in the communication time can be dealt with
in several ways. First, the algorithm can be restructured so that each
processor gets less data at startup. Second, the communication pattern
can be remapped so that the initial data is distributed in multiple
steps along a tree.  Finally, since the initial data is the same for
each processor, it should be possible to use a multicast to allow the data
to be distributed everywhere in a single broadcast.  Without
alleviating this bottleneck the speedup is limited to approximately 100
on a 100 MBit/s class network.  If the initial transmit bottleneck can
be overcome, it should be possible to see speedups in the 5000 range.

\section{Summary and Conclusions}

  We have presented the Adaptive Matched Filter method for the automatic
selection of clusters of galaxies from a galaxy database. The AMF is
adaptive in two ways. First, the AMF uses a two step approach that first
applies a coarse filter to find the clusters and then a fine filter to
provide more precise estimates of the distance and size of each cluster.
Second, the AMF uses the location of the data points as a ``naturally''
adaptive grid to ensure sufficient spatial resolution.

  Matched Filter techniques have a firm mathematical basis in
statistical signal processing.  The AMF uses a hierarchy of two filters
(each mathematically correct under its assumptions).  Combining these
filters allow the AMF to maximize computational performance and
accuracy. The AMF also provides two estimates for each cluster which can
be compared as an additional check.  This is particular effective for
these filters because they react differently when given insufficient
data.

  The AMF relies heavily on models for both the cluster and the
background field.  This prior information is quite extensive and makes
the AMF complex to implement and difficult to adapt to new data sets. 
To alleviate this coding challenge a hybrid coding approach was used to
leverage the ease of use of interpreted languages along with the compute
performance  of compiled languages.  In this way the complex task of
testing model inputs and observing their effect through the data
processing pipeline can be done quickly without sacrificing the compute
efficiency necessary to complete the application in a timely manner.

  A further benefit of the hybrid approach is that it makes available
to the compiled code a wide variety of parallel software libraries and
tools.  A parallel implementation is critical to the application
because matched filter techniques work by testing every possible
location in the cluster space for the presence of a cluster.  This is a
compute intensive operation, but also provides a high degree of
parallelism. The parallelization scheme used for the AMF application is
a client/server approach which is a very effective on
Network-Of-Workstations.  The TNT client/server software used is
lightweight and efficient, and provides a naturally load balancing and
fault tolerant framework.

  The AMF has been extensively tested on simulated data.  These results
indicate that it robustly and accurately detects clusters and estimates
their positions while having few false positives. The AMF is now being
applied to the first results of the Sloan Digital Sky Survey
\cite{Kim00}.  These tests have shown that the AMF detects all
previously known clusters in this data and performs at or above other
cluster finding methods.  The AMF hybrid application architecture has
proven effective in supporting the implementation of new datasets.  The
TNT based client/server parallelization scheme has also demonstrated
significant speedups which will make it feasible for this application
to address to the entire SDSS when it becomes available.

\section*{Acknowledgments}

\begin{figure}[tbh]
\centerline{\includegraphics[width=6.5in]{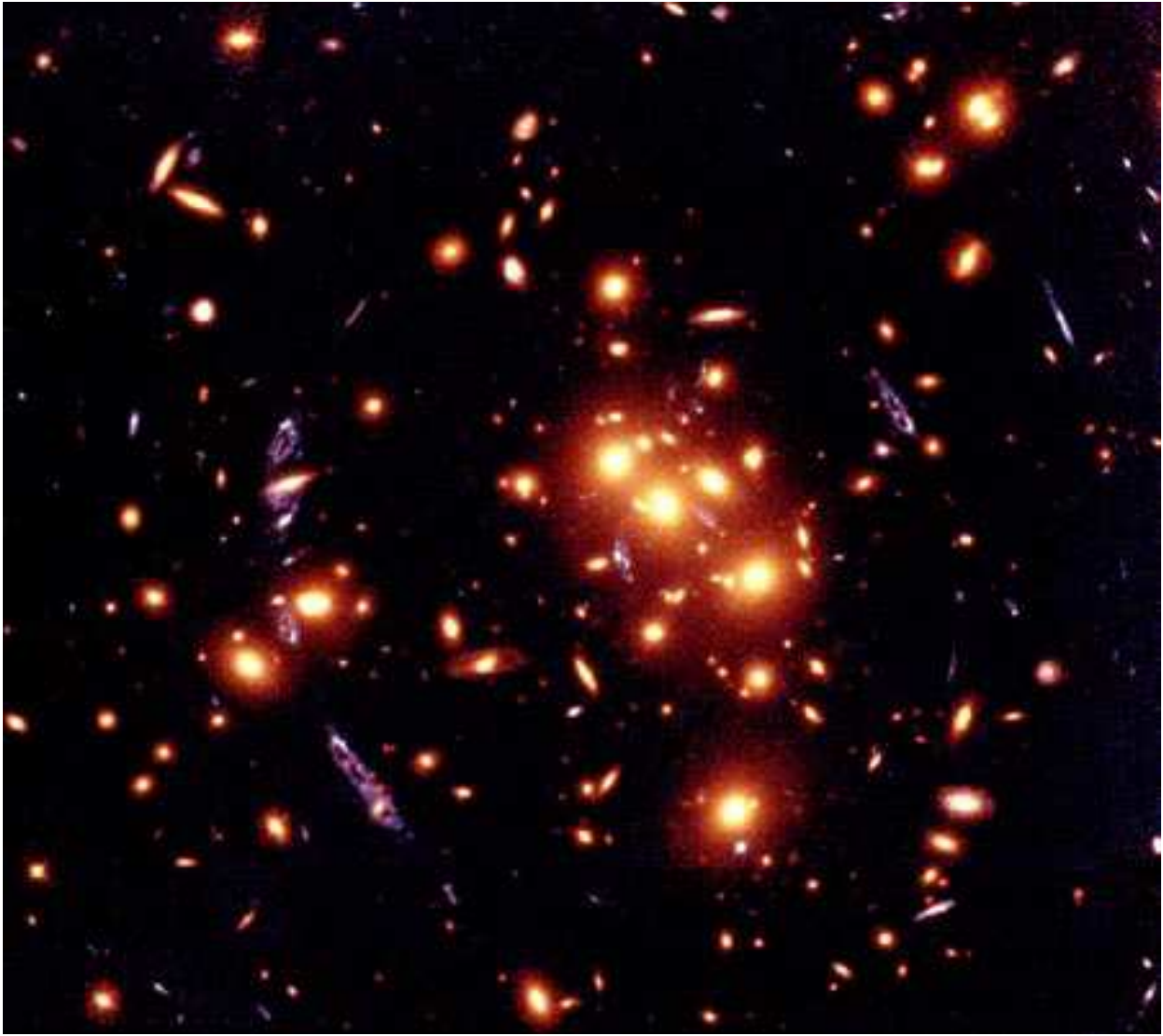}}
\caption{ {\bf Cluster Image.}
  The cluster 0024+1654 as seen with the Hubble Space Telescope
\cite{Colley96}. The reddish circular patches in the center are
Spheroidal galaxies each containing $10^{12}$ stars.  The flatter
reddish patches are Spiral galaxies like our own Milky Way.  The blue
arcs around the edge are from a distant background galaxy that has been
gravitationally "lensed" by the cluster.  This cluster has around one
thousand members and lies at a distance of 4.5 billion light years.
[Note: our nearest neighbor the Andromeda galaxy is 1.5 million light
years away.]
}
\label{fig:colley_cluster}
\end{figure}

\begin{figure}[tbh]
\centerline{\includegraphics[width=6.5in]{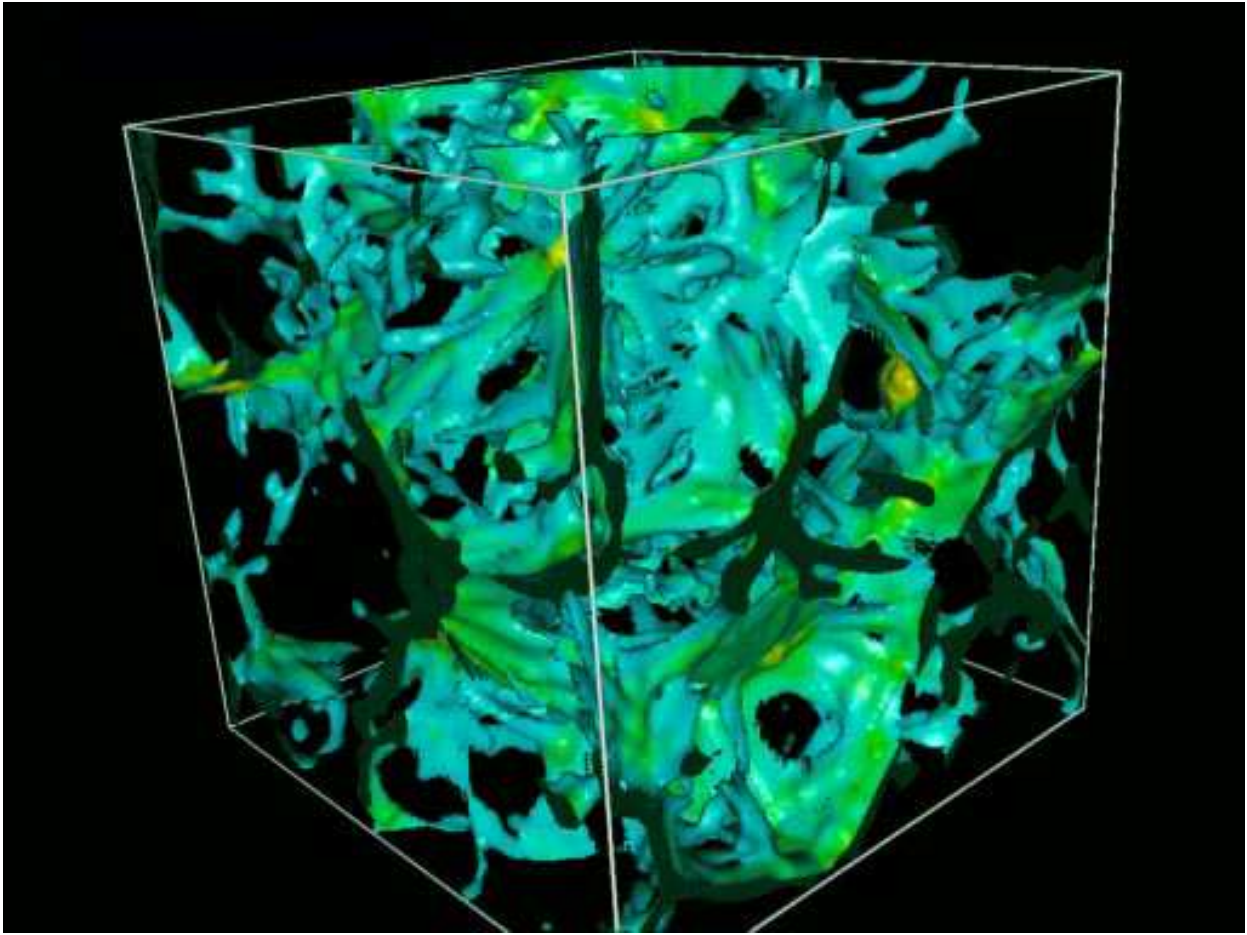}}
\caption{ {\bf Filament Simulation.}
  Isosurface view of a 3D simulation (courtesy of Michael Norman) of the
Universe showing the sheets and filaments of matter along which
galaxies form.  Clusters of galaxies tend to form where two filaments
cross. The simulation volume is approximately 500 light years on a
side.
}
\label{fig:ncsa_sheets}
\end{figure}

Jeremy Kepner and Rita Kim would particularly like to thank their Ph.D.
advisors Profs. David Spergel and Michael Strauss for supporting this
work. The authors are also grateful to Wes Colley and Mike Norman for
providing Figures~\ref{fig:colley_cluster} and \ref{fig:ncsa_sheets}. In
addition, Neta Bahcall, James Gunn, Robert Lupton and David Schlegel
provided invaluable assistance in the development of the AMF, and Aaron
Marks, Maya Gokhale, Ron Minnich, and John Degood were most helpful
in providing the TNT software library.  We are also greatful to Paul
Monticciolo for his helpful comments.


\newpage


\begin{table}[tbh]
\begin{center}
\begin{tabular}{|cccccc|}
\hline
 $N_\CPU$ & $N_\CPU$ (eff) & $N_\task$ & Total Time  & Speedup & Efficiency \\
\hline
   1      &    1.0       &     1     &    5451     &    -      &     -      \\
  13      &   10.5       &    50     &     700     &   7.8     &    74\%    \\
  13      &   10.5       &   200     &     650     &   8.4     &    80\%    \\
  32      &   32.0       &   200     &     200     &  27.3     &    85\%    \\
\hline
\end{tabular}
\end{center}
\caption{ {\bf AMF Execution Times.}
  Execution times in seconds for various numbers of processors and
various numbers of tasks.  $N_\CPU$ is the number of processors used.
$N_\CPU$ (eff) is the number of processors weighted by their clock
speed.  $N_\task$ is how many sub-tasks the problem was broken into. 
The increased parallel efficiency between rows two and three is due to
the use of more tasks which results in less slack time due to the task
granularity.  The increased parallel efficiency between rows three and
four is due to the use of a NOW with a higher performing interconnect
which reduces the time it takes to initially transmit the data.
}
\label{tab:execution_times}
\end{table}


\clearpage
\newpage

\section*{Biographies}
\begin{figure}[tbh]
{\bf Jeremy Kepner} received his B.A. in Astrophysics from Pomona
College (Claremont, CA).  He obtained his Ph.D. focused on Computational
Science from the Dept. of Astrophysics at Princeton University
in 1998, after which he joined MIT Lincoln Lab.  His research
has addressed the development of parallel algorithms and tools
and the application of massively parallel computing to a variety
of data intensive problems. 
E-mail:jvkepner@astro.princeton.edu or kepner@ll.mit.edu
\end{figure}

\begin{figure}[tbh]
{\bf Rita Seung Jung Kim} Rita S.J. Kim  is completing her Ph.D. at
Princeton University in Astrophysical Sciences. She received her B.S. in
Astronomy at Seoul National University (Seoul, Korea), and spent one
year in Paris at the Institut d'Astrophysique de Paris. Her thesis work
concentrates on the properties of cluster galaxies, and has also worked
on statistical modeling of the large scale structure of the Universe.
E-mail:rita@astro.princeton.edu
\end{figure}

\end{document}